\documentclass[nofootinbib,prd,twocolumn,showpacs,showkeys,preprintnumbers]{revtex4-1}
\usepackage{hyperref,amssymb,amsmath,mathrsfs,bm,graphicx}
\usepackage[utf8]{inputenc}
\usepackage{enumitem}
\usepackage{xcolor}
\input{epsf}

\begin{document}
\title {The imprints of the instantaneous  appearance of a conformal Killing vector field on  the evolution of self-gravitating fluid spheres}
\author{L. Herrera}
\email{lherrera@usal.es}
\affiliation{Instituto Universitario de F\'isica
Fundamental y Matem\'aticas, Universidad de Salamanca, Salamanca 37007, Spain.}
\author{A. Di Prisco}
\email{alicia.diprisco@ucv.ve}
\affiliation{Escuela de F\'{\i}sica, Facultad de Ciencias,
Universidad Central de Venezuela, Caracas 1050, Venezuela.}
\author{J. Ospino}\affiliation{Departamento de Matem\'atica Aplicada and Instituto Universitario de F\'isica
Fundamental y Matem\'aticas, Universidad de Salamanca, Salamanca 37007, Spain.}
\email{j.ospino@usal.es}
\date{\today}
\begin{abstract}
We study the influence of the instantaneous  appearance  of a conformal Killing vector (CKV) in self-gravitating  fluid spheres during their evolution.  For doing that we introduce a tensor variable whose time dependence allows the existence of a CKV for a given value of the time-like coordinate. We consider adiabatic and dissipative fluids. The analysis of different relevant physical variables in this process provides a smoking gun signature from the emergence of CKV at some point of the evolution. Prospective applications of these results, as well as open questions and pending issues related to this problem, are discussed.
 \end{abstract}
\date{\today}
\pacs{04.40.-b, 04.40.Nr, 04.40.Dg}
\keywords{Relativistic Fluids, spherical sources, self--similarity, interior solutions.}
\maketitle

\section{Introduction}
The interplay  between physics and symmetry has a long and a venerable history, starting with the seminal paper by Noether \cite{noe}.

Among the different types of symmetry considered in the literature, there is one that has attracted the attention of researchers for many decades. We have in mind the conformal motions, which are characterized by the admittance of a conformal Killing vector (CKV). 

The relevance of conformal motions  in the study of self-gravitating fluids stems from the relevance of self--similar solutions in Newtonian hydrodynamics  \cite{1, 2, 3, 4, 5}, which in turn are related  to the existence of a homothetic Killing vector field (HKV) in general relativity \cite{6}, which imposes specific restrictions on the metric tensor. A natural generalization of a HKV is a CKV.

The great interest aroused by this kind of symmetry is illustrated by the large number of works dedicated to this issue (for a  partial list see   \cite{6, 7, 8co, 9co, 10co, 11co, 12co,13co, 14, 15, 16, 17, 18, 19, 20, 21, 22, 23, 24, 25, 26, 27, 28, 29, 30, 31, 32, 33, 33b, 34, 35, 36, 37, 38, 39, 40, 40bis, 41, 42, 43, 44, sherif, matondo, rej, RS, MH2, HM2, SIF, SHS, Bhar2, TD, DRGR2, OS, ZSRA, DRGR, SNN, BHL1, BHL, RRKKK, K, SIF2, kar, sahoo, uref} and references therein).

In this work we propose to carry on a ``gedanken experiment'' consisting in studying  spherically symmetric fluid distributions which at some point of their  evolution admit a CKV.  The appearance of the CKV is controlled by a tensor defined through a single scalar function, which we name the asymmetry factor,  by means of which we may switch on (off) a CKV. As we will  see later, the suden appearance  of the CKV, even during  an infinitesimal small interval of time,  has a remarkable influence on all physical variables.

To carry on our  ``gedanken experiment'' we shall consider two different models. The first one is adiabatic (non-dissipative), and is obtained under  the assumption of quasi-homologous evolution. In this case the complexity factor is not assumed to have any preassigned specific value.  The evolution of such a system is analyzed in detail  assuming that for  some value of the time-like coordinate, the space-time admits a CKV, which vanishes immediately afterward. Such a CKV is assumed further to be orthogonal to the four-velocity of the fluid. 

The second model is dissipative. In this case the fluid is assumed to be shear-free and the complexity factor vanishes. Also, as in the previous case,  the CKV assumed to appear instantaneously is orthogonal to the four-velocity of the fluid. 

In both cases, all the physical variables are  analyzed in detail. In the adiabatic case particular attention being paid to the effects of the suden appearance of the CKV on the behavior of the complexity factor, whereas in the dissipative case the relevant influence of the appearance of the CKV is focused on the behavior of the pressure and the luminosity.

The manuscript is organized  as follows: the next section is devoted to describe the method to study the evolution  of   self-gravitating dissipative fluid spheres.  In Section III we introduce the concept of asymmetry factor for conformal motions, and express the metric variables in terms of such a factor.
The non-dissipative case is analyzed in section IV, whereas  section V is devoted to describe the dissipative case.  A  discussion on  the obtained results   is  presented in section VI . Finally several appendices are  included containing useful formulae.
\section{Fundamental variables  and relevant equations}
In what follows we shall briefly summarize  definitions and main equations required for describing the structure and evolution of spherically symmetric dissipative fluids. We shall heavily rely on \cite{epjc}, therefore we shall omit many steps in the calculations, details of which the reader may  find in that reference.

We consider  spherically symmetric distributions  of evolving
fluids, which  are  bounded by a spherical surface $\Sigma$. The fluid is
assumed to be locally anisotropic (principal stresses unequal) and undergoing dissipation in the
form of heat flow (diffusion approximation).

Choosing comoving coordinates, the general
interior metric can be written as
\begin{equation}
ds^2=-A^2dt^2+B^2dr^2+R^2(d\theta^2+\sin^2\theta d\phi^2),
\label{1}
\end{equation}
where $A$, $B$ and $R$ are functions of $t$ and $r$ and are assumed
positive. We number the coordinates $x^0=t$, $x^1=r$, $x^2=\theta$
and $x^3=\phi$. Observe that $A$ and $B$ are dimensionless, whereas $R$ has the same dimension as $r$.

The energy--momentum tensor $T_{\alpha\beta}$ of the fluid distribution
has the form
\begin{eqnarray}
T_{\alpha\beta}&=&(\mu +
P_{\perp})V_{\alpha}V_{\beta}+P_{\perp}g_{\alpha\beta}+(P_r-P_{\perp})K_{
\alpha}K_{\beta}\nonumber \\&+&q_{\alpha}V_{\beta}+V_{\alpha}q_{\beta}
, \label{3}
\end{eqnarray}
where $\mu$ is the energy density, $P_r$ the radial pressure,
$P_{\perp}$ the tangential pressure, $q^{\alpha}$ the heat flux, $V^{\alpha}$ the four-velocity of the fluid,
and $K^{\alpha}$ a unit four-vector along the radial direction. These quantities
satisfy
\begin{eqnarray}
V^{\alpha}V_{\alpha}=-1, \;\; V^{\alpha}q_{\alpha}=0, \;\; K^{\alpha}K_{\alpha}=1,\;\;
K^{\alpha}V_{\alpha}=0.
\end{eqnarray}

It will be convenient to express the  energy momentum tensor  (\ref{3})  in the equivalent (canonical) form
\begin{equation}
T_{\alpha \beta} = {\mu} V_\alpha V_\beta + P h_{\alpha \beta} + \Pi_{\alpha \beta} +
q \left(V_\alpha K_\beta + K_\alpha V_\beta\right), \label{Tab}
\end{equation}
with
$$ P=\frac{P_{r}+2P_{\bot}}{3}, \qquad h_{\alpha \beta}=g_{\alpha \beta}+V_\alpha V_\beta,$$

$$\Pi_{\alpha \beta}=\Pi\left(K_\alpha K_\beta - \frac{1}{3} h_{\alpha \beta}\right), \qquad \Pi=P_{r}-P_{\bot}.$$

Since we are considering comoving observers, we have
\begin{eqnarray}
V^{\alpha}&=&A^{-1}\delta_0^{\alpha}, \;\;
q^{\alpha}=qK^{\alpha}, \;\;
K^{\alpha}=B^{-1}\delta^{\alpha}_1,
\end{eqnarray}
where $q$ is a function of $t$ and $r$.

It is worth noticing that we do not explicitly add bulk or shear viscosity to the system because they
can be trivially absorbed into the radial and tangential pressures, $P_r$ and
$P_{\perp}$, of the collapsing fluid (in $\Pi$). Also we do not explicitly  introduce  dissipation in the free streaming approximation since it can be absorbed in $\mu, P_r$ and $q$.

The Einstein equations for (\ref{1}) and (\ref{Tab}), are explicitly written  in Appendix A.

The acceleration $a_{\alpha}$ and the expansion $\Theta$ of the fluid are
given by
\begin{equation}
a_{\alpha}=V_{\alpha ;\beta}V^{\beta}, \;\;
\Theta={V^{\alpha}}_{;\alpha}, \label{4b}
\end{equation}
and its  shear $\sigma_{\alpha\beta}$ by
\begin{equation}
\sigma_{\alpha\beta}=V_{(\alpha
;\beta)}+a_{(\alpha}V_{\beta)}-\frac{1}{3}\Theta h_{\alpha\beta}.
\label{4a}
\end{equation}

From  (\ref{4b}) we have for the  four--acceleration and its scalar $a$,
\begin{equation}
a_\alpha=a K_\alpha, \;\; a=\frac{A^{\prime}}{AB}, \label{5c}
\end{equation}
and for the expansion
\begin{equation}
\Theta=\frac{1}{A}\left(\frac{\dot{B}}{B}+2\frac{\dot{R}}{R}\right),
\label{5c1}
\end{equation}
where the  prime stands for $r$
differentiation and the dot stands for differentiation with respect to $t$.

The shear tensor (\ref{4a}) has only one independent non zero component, and may be written as 
\begin{equation}
\sigma_{\alpha \beta}=\sigma \left(K_\alpha K_\beta-\frac{h_{\alpha \beta}}{3}\right),
\label{defs}
\end{equation}
with 
\begin{equation}
\sigma=\frac{1}{A}\left(\frac{\dot{B}}{B}-\frac{\dot{R}}{R}\right).\label{5b1}
\end{equation}

Next, the mass function $m(t,r)$  reads
\begin{equation}
m=\frac{R^3}{2}{R_{23}}^{23}
=\frac{R}{2}\left[\left(\frac{\dot R}{A}\right)^2-\left(\frac{R^{\prime}}{B}\right)^2+1\right].
 \label{17masa}
\end{equation}

Introducing the proper time derivative $D_T$
given by
\begin{equation}
D_T=\frac{1}{A}\frac{\partial}{\partial t}, \label{16}
\end{equation}
we can define the velocity $U$ of the 
fluid  as the variation of the areal radius with respect to proper time, i.e.
\begin{equation}
U=D_TR , \label{19}
\end{equation}
where $R$ defines the areal radius of a spherical surface inside the fluid distribution (as
measured from its area).

Then (\ref{17masa}) can be rewritten as
\begin{equation}
E \equiv \frac{R^{\prime}}{B}=\left(1+U^2-\frac{2m}{R}\right)^{1/2}.
\label{20x}
\end{equation}
Using (\ref{20x}) we can express (\ref{17a}) as
\begin{equation}
4\pi q=E\left[\frac{1}{3}D_R(\Theta-\sigma)
-\frac{\sigma}{R}\right],\label{21a}
\end{equation}
where   $D_R$ denotes the proper radial derivative,
\begin{equation}
D_R=\frac{1}{R^{\prime}}\frac{\partial}{\partial r}.\label{23a}
\end{equation}
\\

\subsection{The complexity factor}
The complexity factor is a  scalar function intended to describe the degree of complexity of a fluid distribution \cite{ps1,ps2}. It is identified with one of the structure scalars \cite{sc}, more specifically with the scalar defining the trace-free part of the electric Riemann tensor.

Thus, the tensor $Y_{\alpha \beta}$  defining the electric part of the Riemann tensor   given  by 
\begin{equation}
Y_{\alpha \beta}=R_{\alpha \gamma \beta \delta}V^\gamma V^\delta,
\label{electric}
\end{equation}
may be written as 

\begin{eqnarray}
Y_{\alpha\beta}=\frac{1}{3}Y_T h_{\alpha
\beta}+Y_{TF}(K_{\alpha} K_{\beta}-\frac{1}{3}h_{\alpha
\beta})\label{electric'}.
\end{eqnarray}

The scalar $Y_{TF}$ describing the trace-free part of $Y_{\alpha\beta}$ is identified with the complexity factor and may be written as (see \cite{sc, ps1, ps2} for details)
\begin{equation}
Y_{TF}= -8\pi\Pi +\frac{4\pi}{R^3}\int^r_0{R^3\left(D_R {\mu}-3{q}\frac{U}{RE}\right)R^\prime dr}.
\label{Y}
\end{equation}

Thus the scalar $Y_{TF}$ may be expressed  in terms of the anisotropy of pressure, the density inhomogeneity and  the dissipative variables.

In terms of the metric functions the scalar $Y_{TF}$ reads

\begin{eqnarray}
Y_{TF}= \frac{1}{A^2}\left[\frac{\ddot R}{R} - \frac{\ddot B}{B} + \frac{\dot A}{A}\left(\frac{\dot B}{B} - \frac{\dot R}{R}\right)\right]+ \nonumber \\ \frac{1}{ B^2} \left[\frac{A^{\prime\prime}}{A} -\frac{A^{\prime}}{A}\left(\frac{B^{\prime}}{B}+\frac{R^{\prime}}{R}\right)\right] .
\label{itfm}
\end{eqnarray}

\subsection{The exterior spacetime and junction conditions}
When considering  bounded fluid distributions  we have to satisfy the junction (Darmois) conditions if we wish to avoid the presence of thin shells. 

Thus, outside $\Sigma$ we assume that we have the Vaidya
spacetime (i.e.\ we assume all outgoing radiation is massless),
described by
\begin{equation}
ds^2=-\left[1-\frac{2M(v)}{\rho}\right]dv^2-2d\rho dv+\rho^2(d\theta^2
+\sin^2\theta
d\phi^2) \label{1int},
\end{equation}
where $M(v)$  denotes the total mass,
 $v$ is the retarded time, and $\rho$ is a null coordinate.

The matching of the full non--adiabatic sphere  (including viscosity) to
the Vaidya spacetime, on the surface $r=r_{\Sigma}=$ constant, was discussed in
\cite{chan1}.

Thus, from the continuity of the first  differential form it follows (see \cite{chan1} for details), 
\begin{equation}
A dt\stackrel{\Sigma}{=}dv \left(1-\frac{2M(v)}{\rho}\right)\stackrel{\Sigma}{=}d\tau, \label{junction1f}
\end{equation}
\begin{equation}
R\stackrel{\Sigma}{=}\rho(v), \label{junction1f2}
\end{equation}
and 
 \begin{equation}
\left(\frac{dv}{d\tau}\right)^{-2}\stackrel{\Sigma}{=}\left(1-\frac{2m}{\rho}+2\frac{d\rho}{dv}\right), \label{junction1f3}
\end{equation}
where $\tau$ denotes the proper time measured on $\Sigma$.

The continuity of the second differential form produces
\begin{equation}
m(t,r)\stackrel{\Sigma}{=}M(v), \label{junction1}
\end{equation}
and

\begin{eqnarray}
&&2\left(\frac{{\dot R}^{\prime}}{R}-\frac{\dot B}{B}\frac{R^{\prime}}{R}-\frac{\dot R}{R}\frac{A^{\prime}}{A}\right)
\stackrel{\Sigma}{=}-\frac{B}{A}\left[2\frac{\ddot R}{R}
-\left(2\frac{\dot A}{A}
-\frac{\dot R}{R}\right)\frac{\dot R}{R}\right]\nonumber \\&+&\frac{A}{B}\left[\left(2\frac{A^{\prime}}{A}
+\frac{R^{\prime}}{R}\right)\frac{R^{\prime}}{R}-\left(\frac{B}{R}\right)^2\right],
\label{j2}
\end{eqnarray}

where $\stackrel{\Sigma}{=}$ means that both sides of the equation
are evaluated on $\Sigma$ (observe a misprint in eq.(40) in \cite{chan1} and a slight difference in notation).

Comparing (\ref{j2}) with  (\ref{13}) and (\ref{14}) one obtains
\begin{equation}
q\stackrel{\Sigma}{=}P_r.\label{j3}
\end{equation}
Thus   the matching of
(\ref{1})  and (\ref{1int}) on $\Sigma$ implies (\ref{junction1}) and  (\ref{j3}).

Also, we have
\begin{equation}
q\stackrel{\Sigma}{=}\frac{L}{4\pi \rho^2}, \label{20lum}
\end{equation}
where $L_\Sigma$ denotes   the total luminosity of the  sphere as measured on its surface and is given by
\begin{equation}
L \stackrel{\Sigma}{=}L_{\infty}\left(1-\frac{2m}{\rho}+2\frac{d\rho}{dv}\right)^{-1}, \label{14a}
\end{equation}
and where
\begin{equation}
L_{\infty} =-\frac{dM}{dv}\stackrel{\Sigma}{=} -\left[\frac{dm}{dt}\frac{dt}{d\tau}\left(\frac{dv}{d\tau}\right)^{-1}\right],\label{14b}
\end{equation}
is the total luminosity measured by an observer at rest at infinity, which is defined by 
\begin{equation}
L_\infty=4\pi P_{r\Sigma}\left[R^2\left(\frac{R^\prime}{B}+\frac{\dot R}{A}\right)^2 \right]_\Sigma.
\label{lum}
\end{equation}

The boundary redshift $z_\Sigma$ is given by
\begin{equation}
\frac{dv}{d\tau}\stackrel{\Sigma}{=}1+z,
\label{15b}
\end{equation}
with
\begin{equation}
\frac{dv}{d\tau}\stackrel{\Sigma}{=}\left(\frac{R^{\prime}}{B}+\frac{\dot R}{A}\right)^{-1}.
\label{16b}
\end{equation}
Therefore the time of formation of the black hole is given by
\begin{equation}
\left(\frac{R^{\prime}}{B}+\frac{\dot R}{A}\right)\stackrel{\Sigma}{=}E+U\stackrel{\Sigma}{=}0.
\label{17b}
\end{equation}

When Darmois conditions are not satisfied, the boundary surface is a thin shell and we have to assume the Israel conditions \cite{isjc}.

\subsection{The transport equation}
In order to obtain an expression for the temperature we need a transport equation. Assuming a causal dissipative theory (e.g.the Israel-- Stewart theory \cite{19nt, 20nt, 21nt} ) the transport equation for the heat flux reads
\begin{eqnarray}
\tilde \tau h^{\alpha \beta}V^\gamma q_{\beta;\gamma}&+&q^\alpha=-\kappa h^{\alpha \beta}\left(T_{,\beta}+Ta_\beta\right)\nonumber \\&-&\frac{1}{2}\kappa T^2 \left(\frac{\tilde \tau V^\beta}{\kappa T^2}\right)_{;\beta} q^\alpha,
\label{tre}
\end{eqnarray}
where $\kappa$ denotes the thermal conductivity, and $T$ and $\tilde \tau$ denote temperature and relaxation time respectively. 

In the spherically symmetric case under consideration, the transport equation has only one independent component which may be obtained from (\ref{tre}) by contracting with the unit spacelike vector $K^\alpha$, we get
\begin{equation}
\tilde \tau V^\alpha  q_{,\alpha}+q=-\kappa \left(K^\alpha T_{,\alpha}+T a\right)-\frac{1}{2}\kappa T^2\left(\frac{\tilde \tau V^\alpha}{\kappa T^2}\right)_{;\alpha} q.
\label{5}
\end{equation}

Sometimes it is possible to simplify the equation above, in the so called truncated  transport equation, when the last term in (\ref{tre}) may be neglected \cite{t8}, producing 
\begin{equation}
\tilde \tau V^\alpha  q_{,\alpha}+q=-\kappa \left(K^\alpha T_{,\alpha}+T a\right).
\label{5trun}
\end{equation}

\subsection{The homologous and quasi--homologous conditions}

For time dependent systems a full description of the complexity of the fluid requires not only a variable measuring the complexity of the structure of the fluid ($Y_{TF}$), but also we need to describe the complexity of the  pattern of evolution of the system.

In \cite{ps2} the concept of  homologous evolution was introduced, in analogy with the same concept in classical astrophysics, as to represent the simplest mode of evolution of the fluid distribution. 

Thus,    the field equation (\ref{13}) written as
\begin{equation}
D_R\left(\frac{U}{R}\right)=\frac{4 \pi}{E} q+\frac{\sigma}{R},
\label{vel24}
\end{equation}
 can be easily integrated to obtain
\begin{equation}
U=\tilde a(t) R+R\int^r_0\left(\frac{4\pi}{E} q+\frac{\sigma}{R}\right)R^{\prime}dr,
\label{vel25}
\end{equation}
where $\tilde a$ is an integration function, or
\begin{equation}
U=\frac{U_{\Sigma}}{R_{\Sigma}}R-R\int^{r_\Sigma}_r\left(\frac{4\pi}{E} q+\frac{\sigma}{ R}\right)R^{\prime}dr.
\label{vel26}
\end{equation}
If the integral in the above equations vanishes  we have from (\ref{vel25}) or (\ref{vel26}) that
\begin{equation}
 U=\tilde a(t) R \Rightarrow \dot R=A\tilde a R,
 \label{ven6}
 \end{equation}
 where $a$  may be put equal to $1$  without loos of generality by reparametrizying $t$.

Thus we may write the above condition as 
\begin{equation}
\dot R=\alpha A R,
\label{mod7a}
\end{equation}
where $\alpha$ is a unit constant with dimensions of $\frac{1}{length}$.

 This relationship  is characteristic of the homologous evolution in Newtonian hydrodynamics \cite{20n,21n,22n}. In our case, this may occur if the fluid is shear--free and non dissipative, or if the two terms in the integral cancel each other.

In \cite{ps2}, the term  ``homologous evolution'' was used to characterize  relativistic systems satisfying, besides  (\ref{ven6}), the condition
\begin{equation}
\frac{R_I}{R_{II}}=\mbox{constant},
\label{vena}
\end{equation}
where $R_I$ and $R_{II}$ denote the areal radii of two concentric shells ($I,II$) described by $r=r_I={\rm constant}$, and $r=r_{II}={\rm constant}$, respectively.

The important point that we want to stress here is that (\ref{ven6}) does not imply (\ref{vena}).
Indeed, (\ref{ven6})  implies that for the two shells of fluids $I,II$ we have
\begin{equation}
\frac{U_I}{U_{II}}=\frac{A_{II} \dot R_I}{A_I \dot R_{II}}=\frac{R_I}{R_{II}},
\label{ven3}
\end{equation}
that implies (\ref{vena}) only if  $A=A(t)$, which by a simple coordinate transformation becomes $A={\rm constant}$. Thus in the non--relativistic regime, (\ref{vena}) always follows from the condition that  the radial velocity is proportional to the radial distance, whereas in the relativistic regime the condition (\ref{ven6}) implies (\ref{vena}), only if the fluid is geodesic.

In \cite{epjc} the homologous condition was relaxed, leading to what was defined as  quasi--homologous evolution,  restricted only by condition  (\ref{ven6}), implying
\begin{equation}
\frac{4\pi}{R^\prime}B  q+\frac{\sigma}{ R}=0.
\label{ch1}
\end{equation}

\section{The asymmetry factor for conformal motions}
Spacetimes whose line element is defined by (\ref{1}), admitting a CKV, satisfy  the equation
\begin{equation}
\mathcal{L}_\chi g_{\alpha \beta} =2\psi g_{\alpha \beta} \rightarrow \mathcal{L}_\chi g^{\alpha \beta} =-2\psi g^{\alpha \beta}, 
\label{1cmh}
\end{equation}
 where $\mathcal{L}_\chi$ denotes the Lie derivative with respect to the vector field ${\bold \chi}$, which unless specified otherwise, has the general form
\begin{equation}
\bold \chi=\xi(t, r, )\partial_t+\lambda(t, r, )\partial_r,
\label{2cmh}
\end{equation}
 and $\psi$ in principle is a function of $t, r$. The case $\psi=constant$ corresponds to a HKV.
 
 In this work we want to consider fluid distributions which in general do not admit  a CKV, except at a given moment of their evolution.
 
 Thus let us consider 

\begin{equation}
\mathcal{L}_\chi g_{\alpha \beta} -2\psi g_{\alpha \beta} =H_{\alpha \beta},
\label{mod1n}
\end{equation}
 with the additional restriction  $V^\alpha \chi_\alpha=0$,
implying
\begin{equation}
\chi^\rho \partial_\rho g_{\alpha \beta} +g_{\alpha \rho}\partial_\beta \chi^\rho+g_{\beta \rho}\partial_\alpha \chi^\rho-2\psi g_{\alpha \beta}=H_{\alpha \beta},
\label{mod11nn}
\end{equation}
where $H_{\alpha \beta}$ is (so far) an arbitrary symmetric tensor. Obviously  the vector $\chi$ defines a CKV  iff  $H_{\alpha \beta}=0$.

Using (\ref{mod11nn}) one obtains for (\ref{1})
$H_{02}= H_{03}=H_{13}=H_{12}=  H_{23}=0$, and the following equations
\begin{equation}
\psi=\frac{H_{00}}{2A^2}+\frac{\chi^1 A^\prime}{A},
\label{nd1}
\end{equation}
\begin{equation}
\psi=-\frac{H_{22}}{2R^2}+\chi^1\frac{R^\prime}{R}
\label{nd2}
\end{equation}
\begin{equation}
\psi=-\frac{H_{11}}{2B^2}+\chi^1\frac{B^\prime}{B}+\chi^{1\prime},
\label{nd3}
\end{equation}

\begin{equation}
B^2 \dot \chi^1=H_{01}.
\label{nd4}
\end{equation}

In order to proceed further  we need to specify the form of the tensor $H_{\alpha \beta}$. In this work  we have initially  considered  two possible forms of this tensor, namely:
\begin{itemize}
\item 
\begin{equation}
H_{\alpha \beta}=H(t) \chi_\alpha \chi_\beta,
\label{op1}
\end{equation}
\item 
\begin{equation}
H_{\alpha \beta}=H(t) V_\alpha V_\beta,
\label{op2}
\end{equation}
\end{itemize}
where $H$ is a scalar function of $t$, which will be hereafter referred to  as the asymmetry factor. 

However it can be shown that the first option is incompatible with the $QH$ condition which will be used for the non-dissipative case, analyzed in the next Section,  therefore we shall consider only the second option.

In this case it is easy to see that there is only one non vanishing component of the tensor $H_{\alpha \beta}$, namely 
\begin{equation}
H_{00}=\alpha H(t)A^2,
\label{nd11}
\end{equation}
where the asymmetry factor $H(t)$ is dimensionless.

Next, from (\ref{nd4}) it follows that $\dot \chi^1=0$, and from (\ref{nd1}), (\ref{nd2})  and (\ref{nd3}) we obtain
\begin{equation}
\alpha R=B \phi(t),
\label{nd12}
\end{equation}

and
\begin{equation}
A=\alpha Re^{-\frac{\alpha Hr}{2}},
\label{nd13}
\end{equation}
where $\phi$ is arbitrary dimensionless function, and we choose $\chi^1=1$.

Using the results above we shall build up  our two models corresponding to the adiabatic and the dissipative case respectively.
\section{Non dissipative case $q=0$.}
If we assume that the fluid is non-dissipative and impose the $QH$ condition then it follows from (\ref{ch1}) that the fluid is  shear-free, which  implies, using  (\ref{nd12}), that $\phi=b= constant$. 

Feeding back the above into (\ref{13})  (with $q=0$) we obtain after integration
\begin{equation}
R=\frac{b}{\alpha\left[\alpha\int{\Omega(t,r)dr}+f(r)\right]},
\label{nd14}
\end{equation}

\begin{equation}
A=\frac{be^{-\frac{\alpha Hr}{2}}}{\left[\alpha\int{\Omega(t,r)dr}+f(r)\right]},
\label{nd15}
\end{equation}

\begin{equation}
B=\frac{1}{\left[\alpha\int \Omega(t,r)dr+f(r)\right]},
\label{nd16}
\end{equation}
where
\begin{equation}
\Omega(t,r)\equiv \frac{b}{2\alpha} \int\frac{\dot R H}{R^2}dt,
\label{nd17}
\end{equation}
and $f(r)$ is an  arbitrary dimensionless function.

Thus,  the metric is defined up to two functions ($\Omega(t,r), f(r)$).

Then using  the $QH$ condition (\ref{mod7a}) in (\ref{nd17}) we obtain 
\begin{equation}
\dot \Omega  =-\frac{H\alpha}{2}\int{\dot \Omega dr},
\label{nd19b}
\end{equation}
leading to 
\begin{equation}
 \int{\dot \Omega dr} =-be^{-\frac{Hr\alpha}{2}},
\label{nd19}
\end{equation}
and

\begin{equation}
\Omega=\frac{\alpha b}{2}\int H e^{-\frac{H\alpha r}{2}}dt+c(r),
\label{nd20}
\end{equation}
where $c$ is an arbitrary dimensionless function.

Let us assume for $H(t)$ the form
\begin{equation}
H=\alpha(t-t_0),\qquad t_0=constant.
\label{nd21}
\end{equation}

Thus our system evolves without admitting a CKV, until $t=t_0$, when the system admits momentarily a CKV, afterwards the CKV disappears.
\\
Feeding back (\ref{nd21}) into (\ref{nd20})  we obtain
\begin{equation}
\Omega=-\frac{b e^{-\frac{\alpha^2 r(t-t_0)}{2}}}{\alpha r} \left[\alpha(t-t_0)+\frac{2}{\alpha r}\right]+c(r).
\label{nd23}
\end{equation}
From the above we obtain for $ \int{ \Omega dr} $ 

\begin{equation}
\int{ \Omega dr}=\frac{2b}{\alpha ^2 r}e^{-\frac{\alpha^2 r(t-t_0)}{2}} +s(r),
\label{nd24}
\end{equation}
where $s(r)\equiv \int c(r)dr.$

Thus we may write the metric variables as

\begin{equation}
R=\frac{b}{\alpha Z(r,t)},
\label{nd25}
\end{equation}

\begin{equation}
B=\frac{1}{Z(r,t)},
\label{nd26}
\end{equation}

\begin{equation}
A=\frac{\alpha r X(r,t)}{2Z(r,t)},
\label{nd27}
\end{equation}
where $X(r,t)\equiv \frac{2b}{\alpha r}e^{-\frac{\alpha^2 r(t-t_0)}{2}}$, $Z(t,r)\equiv X(t,r)+f(r)$, and the function $f(r)$ now incorporates the function $s(r)$.

Then the field equations read 
\begin{equation}
8\pi \mu=\frac{12 \dot X^2}{\alpha^2 r^2 X^2}+2(X+f)(X^{\prime \prime}+f^{\prime \prime})-3(X^\prime+f^\prime)^2+\frac{\alpha^2}{b^2}(X+f)^2,
\label{mund1}
\end{equation}

\begin{eqnarray}
8\pi P_r&=&-\frac{4(X+f)^2}{\alpha^2 r^2 X^2}\left[-\frac{2\ddot X}{X+f}+\frac{3\dot X^2}{(X+f)^2}+\frac{2\dot X^2}{X(X+f)}\right]\nonumber \\&+&(X+f)^2\left[-2(\frac{1}{r}+\frac{X^\prime}{X})(\frac{X^\prime+f^\prime}{X+f})+\frac{3(X^\prime+f^\prime)^2}{(X+f)^2}\right]\nonumber\\ &-&\frac{\alpha^2 (X+f)^2}{b^2},
\label{mund2}
\end{eqnarray}

\begin{eqnarray}
8\pi P_\bot=-\frac{4(X+f)^2}{\alpha^2 r^2 X^2}\left[-\frac{2\ddot X}{X+f}+\frac{3\dot X^2}{(X+f)^2}+\frac{2\dot X^2}{X(X+f)}\right]\nonumber \\+(X+f)^2\left[\frac{X^{\prime \prime}}{X}-\frac{2(X+f)^{\prime \prime}}{X+f}+\frac{2 X^\prime}{r X} \nonumber \right. \\ \left.-\frac{2(X+f)^\prime}{r(X+f)}+ \frac{3(X^\prime+f^\prime)^2}{(X+f)^2}-\frac{2X^\prime (X+f)^\prime)}{X(X+f)}\right],\nonumber \\,
\label{mund3}
\end{eqnarray}
and for the mass function and the complexity factor we obtain
\begin{equation}
m=\frac{b}{\alpha(X+f)}\left[ \frac{4 b^2 \dot X^2}{\alpha^4 r^2 X^2 (X+f)^2}-\frac{b^2 (X^{\prime}+f^{\prime})^2}{\alpha^2(X+f)^2}+1\right],
\label{mund4}
\end{equation}

\begin{equation}
Y_{TF}=(X+f)^2\left[ \frac{2 X^\prime}{r X} +\frac{X^{\prime \prime}}{X}-\frac{(X+f)^{\prime \prime}}{X+f}\right].
\label{mund5}
\end{equation}

In order to specify a model, let us introduce the dimensionless  variables
\begin{equation}
x\equiv\frac{r}{r_\Sigma},\qquad y\equiv\alpha(t-t_0).
\label{mond1}
\end{equation}
Next, we shall chose for the free parameters the value
\begin{equation}
b= \alpha r_\Sigma=1.
\label{mond2}
\end{equation}
Finally the form of the function $f(r)$ is assumed to be
\begin{equation}
f=\alpha r=x.
\label{mond3}
\end{equation}

From the above it follows at once that
\begin{equation}
X=\frac{2}{x}e^{-\frac{xy}{2}},\qquad Z=\frac{2}{x}e^{-\frac{xy}{2}}+x.
\label{mond4}
\end{equation}

Under the above conditions the physical variables read
\begin{eqnarray}
\frac{8\pi \mu}{\alpha^2}=\frac{e^{-xy}}{x^2}\left(\frac{4}{x^2}-y^2-\frac{4y}{x}+4\right)\nonumber \\+e^{-\frac{xy}{2}}\left(\frac{20}{x^2}+\frac{10y}{x}+y^2+4\right)+x^2,
\label{mund1b}
\end{eqnarray}

\begin{eqnarray}
\frac{8\pi P_r}{\alpha^2}=\frac{e^{-xy}}{x^2}\left(\frac{12}{x^2}+y^2+\frac{8y}{x}-4\right)\nonumber \\-e^{-\frac{xy}{2}}\left(4+y^2+\frac{12}{x^2}+\frac{6y}{x}\right)+x y-x^2,
\label{mund2b}
\end{eqnarray}

\begin{eqnarray}
\frac{8\pi P_\bot}{\alpha^2}&=&-e^{-\frac{xy}{2}}\left(\frac{10y}{x}+\frac{20}{x^2}+y^2\right)-e^{-xy}\frac{4}{x^4}\nonumber \\&+&yx(1+\frac{yx}{4}),
\label{mund3b}
\end{eqnarray}
while the expressions for the complexity factor and the mass function becomes

\begin{eqnarray}
\frac{Y_{TF}}{\alpha^2}=e^{-\frac{xy}{2}}\left(\frac{y^2}{2}-\frac{2y}{x}-\frac{4}{x^2}\right)\nonumber \\-e^{-xy}\left( \frac{8}{x^4}+\frac{4y}{x^3}\right)+\frac{x^2y^2}{4},
\label{mund3b}
\end{eqnarray}
and

\begin{eqnarray}
m \alpha=\frac{1}{(\frac{2}{x}e^{-\frac{xy}{2}}+x)^3}\left[\frac{4}{x^2}e^{-xy}\left(1-\frac{y^2}{4}-\frac{y}{x}-\frac{1}{x^2}\right)\right. \nonumber \\ \left.+\frac{2 e^{-\frac{xy}{2}}}{x}\left(\frac{2}{x}+y+2x\right)+x^2 \right],\nonumber\\
\label{mund4b}
\end{eqnarray}
which evaluated at the boundary surface $x=1$ reads
\begin{eqnarray}
m _\Sigma\alpha=\frac{1}{(2 e^{-\frac{y}{2}}+1)^3}\left[4 e^{-y}\left(-\frac{y^2}{4}-y\right)+2 e^{-\frac{y}{2}}\left(4+y \right)+1\right].\nonumber\\
\label{mund45b}
\end{eqnarray}

At this point the following remarks are in order
\begin{itemize}
\item As it follows from (\ref{nd25}) and (\ref{mond4}), the regularity condition $R(t,0)=0$ is not satisfied. Furthermore it follows from (\ref{mund1b})-(\ref{mund3b}) that physical variables are singular at $r=0$. Accordingly the center of the distribution should be excluded from  consideration.
\item The Darmois conditions are not satisfied on the boundary surface  $r=r_\Sigma$. The radial pressure at the surface never  vanishes.
\item For $y\approx  -4$ the total mass vanishes suggesting  that the star becomes a ghost star. However the radial pressure at the boundary for this value of $y$ does not vanish.
\end{itemize}

In Figure 1, we have plotted the evolution of the complexity factor, the areal radius of the boundary surface, and the anisotropic factor ($P_r-P_\bot)$, which are the physical variables that more clearly illustrate the effect of the momentarily appearance of the CKV at $y=0$. These  results will be discussed further in the last section.

\begin{figure}[h]
\includegraphics[width=5.in,height=6.in,angle=0]{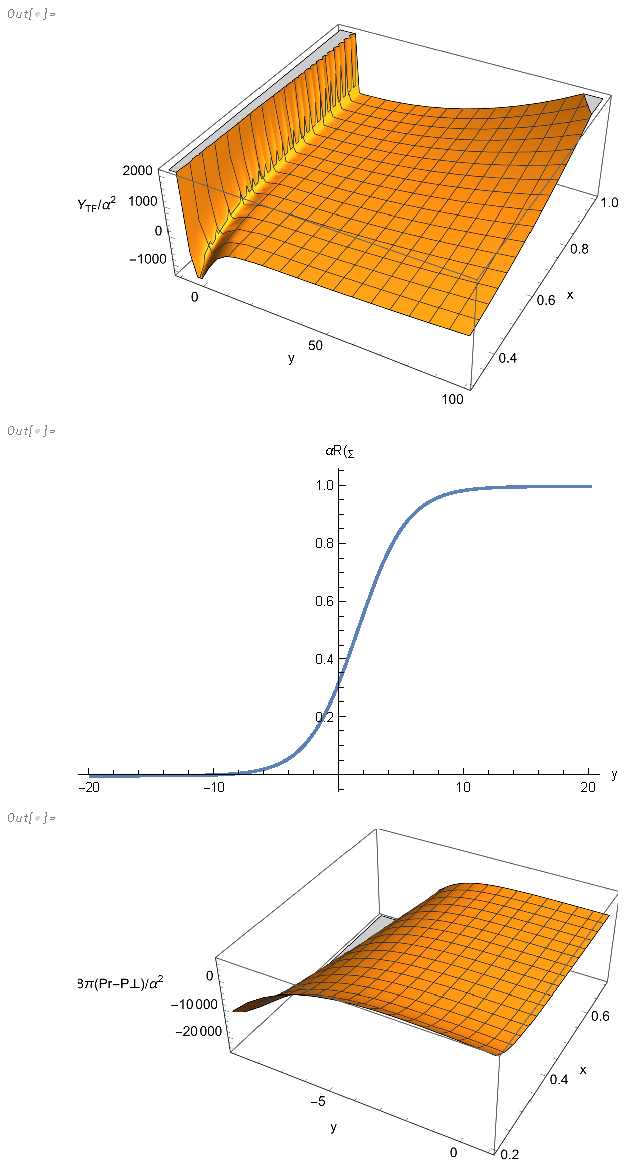}
\caption{$Y_{TF}/\alpha^2$  and $8 \pi(P_r-P_\bot)/\alpha^2$ as functions of $y$ and $x$  in the interval $(y,-10,100)$, $(x, .3, 1)$ and $(y, -9, 1)$ respectively;  $\alpha R_\Sigma$ in the interval $(y, -20, 20)$, for the non-dissipative case}.\label{Figure 1}
\end{figure}

\section{Dissipative case  $q\neq0$}
Let us now consider the dissipative case, for the choice (\ref{op2}) of the tensor $H_{\alpha \beta}$. In such a case relationships (\ref{nd12}) and (\ref{nd13}) still hold.

Thus we have 
\begin{equation}
\alpha R=B\phi(t), \qquad A=\alpha Re^{-\frac{H\alpha r}{2}},
\label{dis1}
\end{equation}
where $\phi(t)$ is a dimensionless function.

In order to specify further the  solutions we shall impose next the  vanishing complexity factor condition ($Y_{TF}=0$) and the shear--free condition.

The latter condition implies that $\phi(t)=constant=b$, whereas the former,  using (\ref{itfm})  leads to the equation 
\begin{equation}
\frac{B^{\prime \prime}}{B}-2\left(\frac{B^\prime}{B}\right)^2+\left(\frac{H\alpha}{2}\right)^2 =0,
\label{dis2}
\end{equation}
whose solution reads
\begin{equation}
B=\frac{f(t)}{\sinh[\frac{H\alpha (r+c_1(t))}{2}]},
\label{dis3}
\end{equation}
leading to 
\begin{equation}
R=\frac{f(t) b}{\alpha\sinh[\frac{H\alpha (r+c_1(t))}{2}]},
\label{dis4}
\end{equation}

\begin{equation}
A=\frac{f(t) b e^{-\frac{H\alpha r}{2}}}{\sinh[\frac{H\alpha (r+c_1(t))}{2}]},
\label{dis5}
\end{equation}
where $f(t)$ is a dimensionless arbitrary function and $c_1(t)$ is an arbitrary function with dimensions of $length$.

From the above  we can write  the physical variables as
\begin{eqnarray}
8\pi \mu &=&\frac{3 e^{2\tilde H}\sinh^2 \tilde H}{b^2 f^2}\left(\frac{\dot f}{f}-\frac{\alpha ^2 r\cosh{\tilde H}}{2 \sinh{\tilde H}}\right)^2\nonumber \\&-&\frac{\alpha^4 (t-t_0)^2  (2+\cosh^2\tilde H)} {4f^2}\nonumber \\&+&\frac{\alpha ^2 \sinh^2 \tilde H}{b^2 f^2},
\label{muidscosf}
\end{eqnarray}

\begin{eqnarray}
4\pi q=\frac{\alpha^2 \sinh^2 \tilde H e^{\tilde H}}{2b f^2}\left\{\frac{\cosh \tilde H}{\sinh \tilde H}[-1+\frac{\dot f (t-t_0)}{f}\right. \nonumber \\ \left. -\frac{\alpha ^2 r(t-t_0)}{2}]+\frac{\dot f (t-t_0)}{f}-\frac{\alpha ^2 r(t-t_0)}{2}\right\},
\label{qdis}
\end{eqnarray}

\begin{eqnarray}
8\pi P_r=-\frac{e^{2\tilde H} \sinh^2\tilde H}{b^2f^2}\left[  \frac{2 \ddot f}{f} -\frac{\dot f^2}{f^2}+\frac{\dot f \alpha^2 r}{f}-\frac{\dot f\alpha^2 r \cosh\tilde H}{f \sinh \tilde H}\right. \nonumber \\ \left. +\frac{\alpha^4 r^2}{4}\left(\frac{2+\cosh^2 \tilde H}{\sinh^2 \tilde H}-\frac{2\cosh \tilde H}{\sinh \tilde H}\right)\right]\nonumber \\ +\frac{\alpha^4(t-t_0)^2\sinh \tilde H \cosh \tilde H}{4f^2}\left(2+\frac{3 \cosh \tilde H}{\sinh \tilde H}\right)-\frac{\alpha ^2 \sinh^2\tilde H}{b^2 f^2} ,\nonumber\\
\label{pridscosf}
\end{eqnarray}

\begin{eqnarray}
8\pi P_\bot=-\frac{e^{2\tilde H} \sinh^2\tilde H}{b^2f^2}\left[  \frac{2 \ddot f}{f} -\frac{\dot f^2}{f^2}+\frac{\dot f \alpha^2 r}{f}-\frac{\dot f\alpha^2 r \cosh\tilde H}{f \sinh \tilde H}\right. \nonumber \\ \left. +\frac{\alpha^4 r^2}{4}\left(\frac{2+\cosh^2 \tilde H}{\sinh^2 \tilde H}-\frac{2\cosh \tilde H}{\sinh \tilde H}\right)\right]\nonumber \\ +\frac{\alpha^4(t-t_0)^2\sinh^2 \tilde H }{4f^2}\left(1+\frac{2 \cosh \tilde H}{\sinh \tilde H}+\frac{2+\cosh^2 \tilde H}{\sinh^2 \tilde H}\right),\nonumber\\
\label{ptidscosf}
\end{eqnarray}
where $\tilde H	\equiv\frac{\alpha r H}{2}$ and we have put $c_1(t)=0$.

In order to specify our model we shall proceed as in the previous case. Thus, let us introduce the dimensionless  variables
\begin{equation}
x\equiv\frac{r}{r_\Sigma},\qquad y\equiv\alpha(t-t_0).
\label{mond1}
\end{equation}
Next, we shall chose for the free parameters the value
\begin{equation}
b= \alpha r_\Sigma=1,
\label{mond2}
\end{equation}
and $H=\alpha(t-t_0)=y$.

From the above it follows that $\tilde H=\frac{xy}{2}$ and $\dot f=f_{,y}\alpha$, $\ddot f=f_{,yy}\alpha^2$, where the subscript$(,y)$ denote derivative with respect to $y$.

Finally, for completely determining the model we need to specify the function $f$. In principle it could be obtained  from the junction condition (\ref{j3}). Hoverer the resulting equation is impossible to solve analytically  and therefore we shall assume the simplest expression ensuring a regular behavior for the metric  and the physical variables (in particular at $y=0$), namely
\begin{equation}
f(t)=y.
\label{deff}
\end{equation}

Under these conditions the physical variables read
\begin{eqnarray}
8\pi \mu/\alpha^2 &=&\frac{3 e^{xy}\sinh^2 (xy/2)}{y^2}\left(\frac{1}{y}-\frac{x \cosh(xy/2)}{2 \sinh(xy/2)}\right)^2\nonumber \\ &-&\frac{(2+\cosh^2(xy/2))}{4}+\frac{\sinh^2(xy/2)}{y^2},
\label{muidscosf}
\end{eqnarray}

\begin{eqnarray}
8\pi q/\alpha^2=\frac{\sinh^2 (xy/2)e^{xy/2}}{y^2}\left\{-\frac{\cosh (xy/2)}{\sinh(xy/2)} \frac{ xy}{2}+1-\frac{xy}{2}\right\},\nonumber \\\label{qdisb}
\end{eqnarray}

\begin{eqnarray}
8\pi P_r/\alpha^2=-\frac{e^{xy} \sinh^2(xy/2)}{y^2}\left[  -\frac{1}{y^2}+\frac{x}{y}\right. \nonumber \\ \left. -\frac{x \cosh{(xy/2)}}{y\sinh(xy/2)}+\frac{ x^2}{4}\left(\frac{2+\cosh^2 (xy/2)}{\sinh^2 (xy/2)}-\frac{2\cosh (xy/2)}{\sinh (xy/2)}\right)\right]\nonumber \\ +\frac{\sinh (xy/2)\cosh (xy/2)}{4}\left(2+\frac{3 \cosh (xy/2)}{\sinh (xy/2)}\right)\nonumber \\-\frac{\sinh^2 (xy/2)}{y^2} ,\nonumber \\
\label{pridscosfb}
\end{eqnarray}

\begin{eqnarray}
8\pi P_\bot/\alpha^2=-\frac{e^{xy} \sinh^2(xy/2)}{y^2}\left[ -\frac{1}{y^2}+\frac{x}{y} \right. \nonumber \\ \left. -\frac{x \cosh{(xy/2)}}{y\sinh(xy/2)}+\frac{ x^2}{4}\left(\frac{2+\cosh^2 (xy/2)}{\sinh^2 (xy/2)}-\frac{2\cosh (xy/2)}{\sinh (xy/2)}\right)\right]\nonumber \\+\frac{\sinh^2(xy/2)}{4}\left(1+\frac{2 \cosh(xy/2)}{\sinh(xy/2)}+\frac{2+\cosh^2(xy/2)}{\sinh^2(xy/2)}\right).\nonumber \\
\label{ptdscosfb}
\end{eqnarray}

Finally from (\ref{14a}) and (\ref{lum}) we obtain 
\begin{equation}
L_\Sigma=\frac{e^{y/2}}{2}\left(1-\frac{y}{2}-\frac{y \cosh (y/2)}{2 \sinh(y/2)}\right),
\label{lums}
\end{equation}
and
\begin{eqnarray}
L_\infty=L_\Sigma\left[\frac{e^y}{y^2}\left(1 -\frac{y  \cosh(y/2)}{2\sinh(y/2)}\right)^2   \nonumber \right. \\  \left.-\frac{\cosh(y/2)e^{y/2}}{\sinh(y/2)}\left(1-\frac{y\cosh(y/2)}{2\sinh(y/2)}\right)+\frac{y^2 \cosh^2(y/2)}{4 \sinh^2(y/2)}\right],\nonumber \\
\label{lumsin} 
\end{eqnarray}
whereas the time for the eventual black hole formation (\ref{17b}) comes from the equation
\begin{equation}
-\frac{y\cosh(y/2)}{2\sinh(y/2)}+\frac{e^{y/2}}{y}\left(1-\frac{y \cosh(y/2)}{2\sinh(y/2)}\right)=0.
\label{bhf}
\end{equation}

Using the truncated version of the transport equation (\ref{5trun}), and (\ref{qdisb}) we may obtain an expression for the temperature. Unfortunately though such  an expression contains hypergeometric functions  making it impossible to obtain any useful information without resorting to numerical analysis. Therefore  we will dispense with this expression and the corresponding graphic.

To highlight the effects derived from the appearance of the CKV on the evolution of the fluid we have plotted $L_\Sigma$, the areal radius of the boundary surface and the condition for the black hole formation in Figure 2, whereas the radial and tangential pressure are plotted in Figure 3.

All these results together with the ones obtained from the previous case will be discussed next.

\begin{figure}[h]
\includegraphics[width=5.in,height=6.in,angle=0]{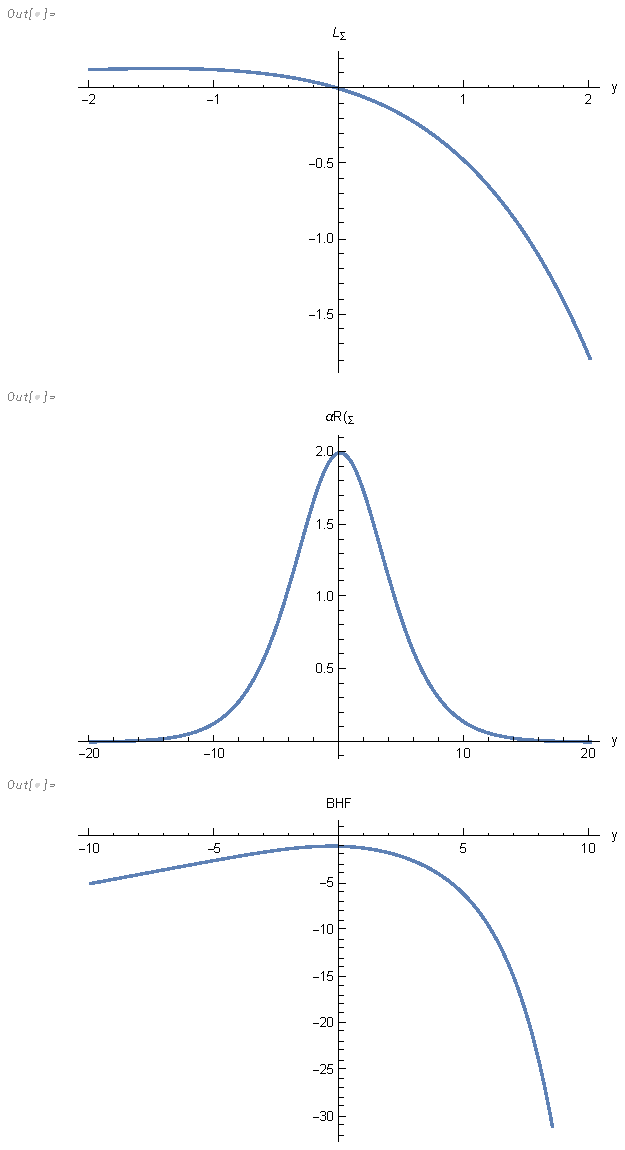}
\caption{$L_\Sigma$, $\alpha R_\Sigma$  and $BHF$  in the interval $(y, -2, 2)$, $(y, -20, 20)$ and $(y, -10, 10)$ respectively, for the dissipative case}.\label{Figure 2}
\end{figure}
.
\begin{figure}[h]
\includegraphics[width=5.in,height=6.in,angle=0]{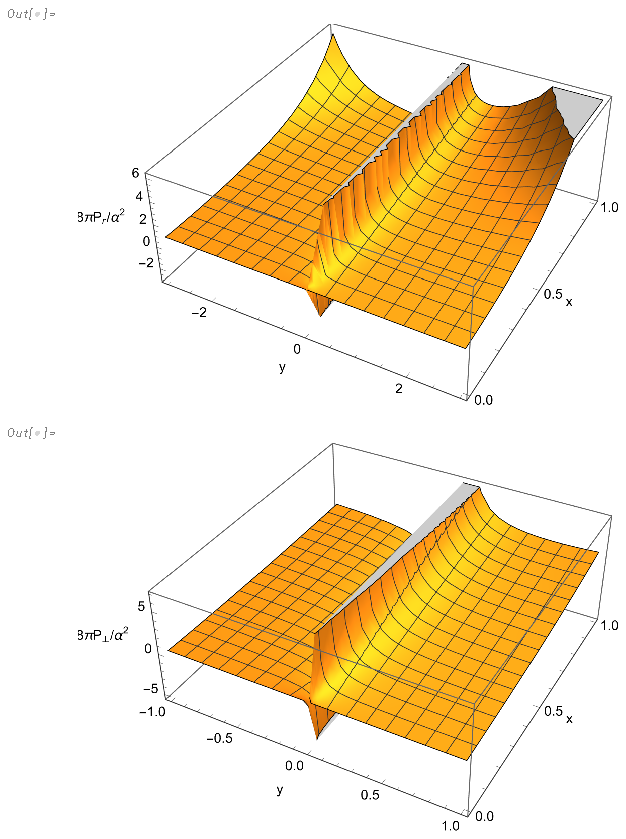}
\caption{$8 \pi P_r/\alpha^2$ and $P_\bot/\alpha^2$ as functions of $y$ and $x$  in the interval $(y, -3, 3)$, $(x, 0, 1)$ and $(y, -1, 1)$, $(x, 0, 1)$ respectively;  for the dissipative case}.\label{Figure 3}
\end{figure}

\section{Discussion}
The purpose of this work is to bring out the effects of an instantaneous appearance of a CKV,  on the evolution of relativistic self-gravitating fluids. For doing that we have introduced the concept of asymmetry factor which allows us to turn on (off) a CKV at any given point of the evolution. A general formalism derived from the above mentioned concept was then applied to study two different models. At this point it is worth emphasizing that these models are not intended to  describe any specific astrophysical scenario,  but are presented with the purpose mentioned above.

The first model was assumed to be non-dissipative. It was obtained under the assumption of quasi-homologous evolution, and the CKV was assumed to be orthogonal to the four-velocity.
Further restrictions on this model concern the specific choice of the tensor $H_{\alpha \beta}$. We have chosen the option (\ref{op2}), since  option (\ref{op1}) together with $QH$ condition leads to 
 $HB^2$ being a function of $r$  alone. However we are interested in situations where $H(t)$ is zero at some point of the evolution, and since $B$ should be regular and different from zero at all times, we should have $H=0$ for any value of $t$. In other words the assumption of $QH$ regime is incompatible with option (\ref{op1}) within the context of this work.

Of course it is possible that imposing restrictions different from the $QH$ regime, one could find a model compatible with  option (\ref{op1}).

Under the conditions above we obtain the model described by (\ref{mund1b})--(\ref{mund45b}). It represents an expanding fluid sphere which admit instantaneously a CKV at $y=0$.
Darmois conditions are not satisfied on the boundary surface of the fluid distribution, accordingly we assume that the boundary is a  thin shell. 

The effects of the appearance of a CKV at  $y=0$ are of two kinds. On the one hand, as depicted in the graphic for the complexity factor in Figure 1, this variable decreases sharply  as we approach $y=0$, increasing afterward  but never returning to the past values. Similar situation appears in the behavior of the anisotropic factor as illustrated in Figure 1. In other words there appears a kind of ``hysteresis''  (memory) effect. The complexity factor ``remembers''  the appearance of the CKV,  after the latter has been turned off. This is remarkable, since it describes for the first time (to our knowledge) a direct relationship  between the complexity factor  and  symmetry (a conformal motion in this particular case).

On the other hand the instantaneous appearance of a CKV may have a direct influence on the evolution of the fluid, not related to any ``memory effect''. Thus the graphic of the evolution of the areal radius of the boundary in Figure 1, shows the sharp increasing in the growth of this variable as $y\rightarrow0$,  returning afterward  to the same pace as before.

The second model is dissipative, it was obtained for the same choice of tensor $H_{\alpha \beta}$ as in the first model and the CKV is assumed also to be orthogonal to the four-velocity. Besides, this model satisfies the vanishing complexity factor condition and is shear-free. The full description of the model is provided by equations (\ref{muidscosf})--(\ref{bhf}). As in the previous example, Darmois conditions are not satisfied on the boundary surface (the condition (\ref{j3}) is only satisfied for $y\approx -0.9$).

As depicted in the graphic for the areal radius of the boundary surface, in Figure 2, this model represents an initially expanding sphere reaching a maximum value of $R_\Sigma$ at $y=0$ (when the the CKV appears) becoming a contracting sphere afterward.  The graphic of the luminosity at the boundary surface indicates that it vanishes at $y=0$ changing from positive to negative for $y>0$. We omit the graphic for the luminosity measured by an observer at infinity since it is  very similar to the latter one.  The last graphic of Figure 2  shows that the condition for the formation of black hole is never satisfied. 

An interesting effect on both pressures is depicted in Figure 3. Indeed at $y=0$ they experience a strong discontinuity, which occurs for any value of the radial coordinate.

Thus as it follows from the above comments the instantaneously appearance of a CKV produces important effects on the subsequent evolution of the system, some of which are directly related to observable variables (e.g luminosity and the gravitational redshift). However two questions are in order at this point, which deserve to be discussed:
\begin{itemize}
\item One wonders to what extent  are the results exhibited in this work model independent?
\item What could be the applications of such effects in the study of self-gravitating systems?
\end{itemize}  

Regarding the first question, it is obvious that some specific effects related to the evolution of the fluid after the appearance of the CKV are tightly depending on the model itself. This is particularly true in what concerns the effect of the second kind mentioned above (not related to any ``hysteresis type phenomenon''. Instead, ``memory'' effects of any kind seem to be intrinsically related to the instantaneous vanishing of the asymmetry factor in general. A more precise answer to this question would require more research work, probably involving numerical methods. 

Concerning the second question, we contemplate the possibility to explain some observational astrophysical or cosmological data, which appear ininteligible in terms of the standard theoretical setup, by resorting to a general approach based on  the asymmetry factor (not necessarily associated to a CKV). As vague as this answer might look like, we believe that this is, right now, the more promising  application of our method.

Finally we would like to conclude mentioning some pending issues that we think deserve to be investigated further 
\begin{itemize}
\item We were unable to make an evaluation of the temperature, based exclusively on analytical methods. However it  is a very important variable, and so it would be interesting to extend the research by resorting to numerical methods to elucidate this issue.
\item We have considered the asymmetry factor related to conformal motions (CKV). However it is obvious that the same approach used here could be extended to other types of symmetry (e.g.  isometries, affine collineations, Ricci collineations, curvature collineations, matter collineations, etc).
\item In our study we have adopted the specific form of the tensor $H_{\alpha \beta}$ given by (\ref{op2}). However, it is likely that more general forms of such a tensor could lead to interesting results not appearing in this work, in particular it could imply  the existence of more than one asymmetry factor . 
\item It is of utmost relevance  to find out if there is any physical phenomenon behind the appearance (disappearance) of any specific symmetry. In other words, can we identify a physical reason for the vanishing of the asymmetry factor?
\end{itemize}

\begin{acknowledgments} 
This  work  was partially supported by the Grant PID2024-158938NB-I00 funded by MICIU/AEI/ 10.13039/501100011033 and by ERDF A way of making Europe.
\end{acknowledgments}
\appendix
\section{Einstein equations}
 Einstein's field equations for the interior spacetime (\ref{1}) are given by
\begin{equation}
G_{\alpha\beta}=8\pi T_{\alpha\beta},
\label{2}
\end{equation}
and its non zero components
with (\ref{1}) and (\ref{3}) 
become

\begin{eqnarray}
8\pi T_{00}&=&8\pi  \mu A^2
=\left(2\frac{\dot{B}}{B}+\frac{\dot{R}}{R}\right)\frac{\dot{R}}{R}
\nonumber \\&-&\left(\frac{A}{B}\right)^2\left[2\frac{R^{\prime\prime}}{R}+\left(\frac{R^{\prime}}{R}\right)^2
-2\frac{B^{\prime}}{B}\frac{R^{\prime}}{R}-\left(\frac{B}{R}\right)^2\right],\nonumber \\
\label{12} 
\end{eqnarray}

\begin{eqnarray}
8\pi T_{01}=-8\pi qAB
=-2\left(\frac{{\dot R}^{\prime}}{R}
-\frac{\dot B}{B}\frac{R^{\prime}}{R}-\frac{\dot
R}{R}\frac{A^{\prime}}{A}\right),\nonumber \\
\label{13} 
\end{eqnarray}
\begin{eqnarray}
8\pi T_{11}&=&8\pi P_r B^2
=-\left(\frac{B}{A}\right)^2\left[2\frac{\ddot{R}}{R}-\left(2\frac{\dot A}{A}-\frac{\dot{R}}{R}\right)
\frac{\dot R}{R}\right]
\nonumber \\&+&\left(2\frac{A^{\prime}}{A}+\frac{R^{\prime}}{R}\right)\frac{R^{\prime}}{R}-\left(\frac{B}{R}\right)^2,
\label{14} 
\end{eqnarray}
\begin{eqnarray}
8\pi T_{22}&=&\frac{8\pi}{\sin^2\theta}T_{33}=8\pi P_{\perp}R^2
=\nonumber \\&-&\left(\frac{R}{A}\right)^2\left[\frac{\ddot{B}}{B}+\frac{\ddot{R}}{R}
-\frac{\dot{A}}{A}\left(\frac{\dot{B}}{B}+\frac{\dot{R}}{R}\right)
+\frac{\dot{B}}{B}\frac{\dot{R}}{R}\right]\nonumber \\
&+&\left(\frac{R}{B}\right)^2\left[\frac{A^{\prime\prime}}{A}
+\frac{R^{\prime\prime}}{R}-\frac{A^{\prime}}{A}\frac{B^{\prime}}{B}
+\left(\frac{A^{\prime}}{A}-\frac{B^{\prime}}{B}\right)\frac{R^{\prime}}{R}\right].\nonumber \\ \label{15}
\end{eqnarray}

The component (\ref{13}) can be rewritten with (\ref{5c1}) and
(\ref{5b1}) as
\begin{equation}
4\pi qB=\frac{1}{3}(\Theta-\sigma)^{\prime}
-\sigma\frac{R^{\prime}}{R}.\label{17a}
\end{equation}
\section{Intermediate calculations for the non-dissipative model}
\begin{eqnarray}
\frac{\dot X}{X}&=&\frac{\alpha^2 r}{2}, \frac{\ddot X}{X}=\frac{\alpha^4 r^2}{4}, \frac{\dot X^\prime}{X}=\frac{\alpha^4 r(t-t_0)}{4},\nonumber \\ \frac{X^\prime}{X}&=&-\left[\frac{1}{r}+\frac{\alpha^2 (t-t_0)}{2}\right], \nonumber \\
\frac{X^{\prime \prime}}{X}&=&\frac{2}{r^2}+\frac{\alpha^2 (t-t_0)}{r}+\frac{\alpha^4 (t-t_0)^2}{4}.
\label{and2}
\end{eqnarray}


\begin{thebibliography}{100}
\bibitem{noe}Noether, E. Invariante Variationsprobleme. {\em Nachrichten von der Gesellschaft der Wissenschaften zu Gottingen. Mathematisch-Physikalische Klasse}, {\bf 1918},  235-257.
\bibitem{1} Barenblatt, G.I.; Zeldovich, Ya.B.  Self-Similar Solutions as Intermediate Asymptotics. {\em Ann. Rev. Fluid. Mech.} {\bf 1972}, {\em  4}, 285--312.
\bibitem{2} Sedov, L. I. Propagation of strong shock waves. {\em J. Appl. Math. Mech.} {\bf 1946}, {\em 10}, 241--250.
\bibitem{3} Sedov, L. I. \textit{Similarity and Dimensional Methods in Mechanics}; Academic. New York, USA, 1967.
\bibitem{4} Taylor, G.I. The Formation of a Blast Wave by a Very Intense Explosion. II. The Atomic Explosion of 1945. {\em Proc. Roy. Soc.}  {\bf 1950}, {\em 201}, 175--186.
\bibitem{5}  Zeldovich, Ya.B.; Raizer, Yu.P.  \textit{Physics of Shock Waves and High Temperature}; Academic. New York, USA, 1963.
\bibitem{6} Cahill, M.E.; Taub, A.H. Spherically symmetric similarity solutions of the Einstein field equations for a perfect fluid. {\em Commun. Math. Phys.} {\bf 1971}, {\em 21}, 1--40.
\bibitem{7}Herrera, L.; Jimenez, J.; Leal, L.; Ponce de Leon, J.; Esculpi, M.; Galina, V. Anisotropic fluids and conformal motions in general relativity. {\em J. Math. Phys.} {\bf 1984}, {\em 25}, 3274--3278.
\bibitem{8co}Herrera,L.; Ponce de Leon, J. Isotropic spheres admitting a one parameter group of conformal motions. {\em J. Math. Phys.} {\bf 1985}, {\em 26}, 778--784.
\bibitem{9co}Herrera,L.; Ponce de Leon, J. Anisotropic  spheres admitting a one parameter group of conformal motions. {\em J. Math. Phys.} {\bf 1985}, {\em 26}, 2018--2023.
\bibitem{10co}Herrera,L.; Ponce de Leon, J. Isotropic and anisotropic charged  spheres admitting a one parameter group of conformal motions. {\em J. Math. Phys.} {\bf 1985}, {\em 26}, 2302--2307.
\bibitem{11co}Herrera,L.; Ponce de Leon, J. Confined gravitational fields produced by anisotropic spheres. {\em J. Math. Phys.} {\bf 1985}, {\em 26}, 2847--2849.

\bibitem{12co}Maartens, R.; Mason, P.S.; Tsamparlis, M. Kinematic and dynamic properties of conformal Killing vectors in anisotropic fluids. {\em J. Math. Phys.} {\bf 1986}, {\em 27}, 2987--2994.

\bibitem{13co}Duggal, K.L.; Sharma, R. Conformal collineations and anisotropic fluids in general relativity. {\em J. Math. Phys.} {\bf 1986}, {\em 27}, 2511--2513.

\bibitem{14}Esculpi, M.; Herrera, L. Conformally symmetric radiating spheres in general relativity. {\em J. Math. Phys.} {\bf 1986}, {\em 27}, 2087--2096.


\bibitem{15}Duggal, K.L. Relativistic fluids with shear and timelike conformal collineations. {\em J. Math. Phys.} {\bf 1987}, {\em 28}, 2700--2704.

\bibitem{16}Mason, D.P.; Maartens, R. Kinematics and dynamics of conformal collineations in relativity. {\em J. Math. Phys.} {\bf 1987}, {\em 28}, 2182--2186.

\bibitem{17}Di Prisco, A.; Herrera,  L.; Jimenez, J.; Galina , V.;  Ibanez, J. The Bondi metric and conformal motions. {\em J. Math. Phys.} {\bf 1987}, {\em 28}, 2692--2696.

\bibitem{18}Duggal, K.L. Relativistic fluids and metric symmetries. {\em J. Math. Phys.} {\bf 1989}, {\em 30}, 1316--1322.

\bibitem{19}A.A. Coley.; B.O.J. Tupper. Special conformal Killing vector space-times and symmetry inheritance {\em J. Math. Phys.}, {\bf 1989} {\em 30}, 261--2625.

\bibitem{20}A.A. Coley.; B.O.J. Tupper. Spacetimes admitting inheriting conformal Killing vector fields  {\em Classical Quantum Grav.} {\bf 1990}, {\em 7}, 1961--1981.

\bibitem{21}A.A. Coley.; B.O.J. Tupper. Spherically symmetric spacetimes admitting inheriting conformal Killing vector fields.  {\em Classical Quantum Grav.} {\bf 1990}, {\em 7}, 2195--2214.

\bibitem{22}Maartens, R; Maharaj, M.S. Conformally symmetric static fluid spheres. {\em J. Math. Phys.} {\bf 1990}, {\em 31}, 151--155.
\bibitem{23}Di Prisco, A.; Herrera, L.; Esculpi, M.  Self-similar scalar soliton star in the thin wall approximation. {\em Phys. Rev. D} {\bf 1991}, {\em 44}, 2286--2294.
\bibitem{24} Saridakis, E.; Tsamparlis, M. Symmetry inheritance of conformal Killing vectors. {\em J. Math. Phys.} {\bf 1991}, {\em 32}, 1541--1551.
\bibitem{25}Aguirregabiria, J.M.; Di Prisco, A.; Herrera, L.; Ibanez, J. Time evolution of self--similar scalar soliton stars: A general study. {\em Phys. Rev. D} {\bf 1992}, {\em 46}, 2723--2725. 
\bibitem{26}Maartens, R.;Maharaj, S.D.; Tupper, B.O.J. General solution and classification of conformal motions in static spherical spacetimes.  {\em Classical Quantum Grav.} {\bf 1995}, {\em 12}, 2577--2586.
\bibitem{27} Maharaj, S.D.; Maartens, R.; Maharaj, M.S.  Conformal symmetries in static spherically symmetric spacetimes.  {\em Int. J. Theor. Phys.} {\bf 1995}, {\em 34}, 2285--222901.
\bibitem{28}Carot , J.;  Sintes,  A. Homothetic perfect fluid spacetimes. {\em Classical Quantum Grav.} {\bf 1997}, {\em 14},  1183--1205. 
\bibitem{29} Carr, B.J.;  Coley, A. A. TOPICAL REVIEW: Self-similarity in general relativity. {\em Classical Quantum Grav.} {\bf 1999}, {\em 16}, R31--R71. 
\bibitem{30}  Barreto, W.;  da Silva, A.  Self-similar and charged spheres in the diffusion approximation. {\em Classical Quantum Grav.} {\bf 1999}, {\em 16}, 1783--1792.
\bibitem{31}  Yavuz, I. ; Yilmaz, I;   Baysal, H. Strange Quark Matter Attached to the String Cloud in the Spherical Symmetric Space-Time Admitting Conformal Motion. {\em Int. J. Mod. Phys. D}  {\bf 2005}, {\em 14}, 1365--1372 . 
\bibitem{32}  Sharif, M.;  Sheikh, U. Timelike and Spacelike Matter Inheritance Vectors in Specific Forms of Energy-Momentum Tensor. {\em Int. J. Mod. Phys. A} {\bf 2006}, {\em 21}, 3213--3234.
\bibitem{33} Barreto, W.; Rodriguez, B.; Rosales, L.; Serrano, O. Self--similar and charged radiating spheres: an anisotropic approach. {\em Gen. Relativ. Gravit.} {\bf 2007}, {\em 39}, 23--39.
\bibitem{33b}Mak, M.K.; Harko, T. Quark stars admitting a one parameter group of conformal motions. {\em Int. J. Mod. Phys. D.} {\bf 2004}, {\em 13}, 149--156.
\bibitem{34} Moopanar, S.; Maharaj, S.D. Conformal symmetries of spherical spacetimes. {\em Int. J. Theor. Phys.} {\bf 2010}, {\em 49}, 1878--1885.
\bibitem{35} Bhar, P. Vaydya--Tikekar--type superdense star admitting conformal motion in presence of quintessence field. {\em Eur. Phys. J. C} {\bf 2015}, {\em 75}, 123.
\bibitem{36} Apostolopoulos, P.S. Spatially inhomogeneous and irrotational geometries admitting intrinsic conformal symmetries. {\em Phys. Rev. D} {\bf 2016}, {\em 94}, 124052.
\bibitem{37}Shee, D.; Rahaman, F.; Guha, B.K.; Ray, S. Anisotropic stars with non--static conformal symmetry. {\em Astr. Space Sci.} {\bf 2016}, {\em 361}, 167.
\bibitem{38}Majonjo, A.; Maharaj, S.D.; Moopanar, S. Conformal vectors and stellar models. {\em Eur. Phys. J. Plus} {\bf 2017}, {\em 132}, 62.
\bibitem{39} Newton Singh, K.; Murad, M.; Pant, N. A 4D spacetime embedded in a 5D pseudo--Euclidean space describing interior compact stars. {\em Eur. Phys. J. A} {\bf 2017}, {\em 53}, 21.
\bibitem{40}Shee, D.; Deb, D.; Ghosh, S.; Guha, B.K.; Ray, S. On the features of Matese--Whitman mass function. {\em arXiv: 1706.00674} {\bf 2017}.
\bibitem{40bis}Herrera, L.; Di Prisco, A. Self--similarity  in static axially symmetric relativistic fluid. {\em Int. J. Mod. Phys. D} {\bf 2018}, {\em 27}, 1750176.
\bibitem{41} Ojako, S.; Goswami, R.; Maharaj, S.D. New class of solutions in conformally symmetric massless scalar field collapse. {\em Gen. Relativ. Gravit.} {\bf 2021}, {\em 53}, 13.
\bibitem{42} Shobhane, P.; Deo, S. Spherically symmetric distributions of wet dark fluid admitting conformal motions. {\em Adv. Appl. Math. Sci.} {\bf 2021},  {\em 20}, 1591--1598.
\bibitem{43} Jape,  J.;  Maharaj, S.D.; Sunzu, J.; Mkenyeleye, J. Generalized compact star models with conformal symmetry. {\em Eur. Phys. J. C } {\bf 2021}, {\em 81}, 2150121.
\bibitem{44} Ivanov,  B. Generating solutions for charged stellar models in general relativity. {\em Eur. Phys. J. C } {\bf 2021}, {\em 81}, 227.
\bibitem{sherif} Sherif, A.;  Dunsby, P.; Goswami , R.; Maharaj, S.D. On homothetic Killing vectors in stationary axisymmetric vacuum spacetimes. {\em Int. J. Geom. Meth. Mod. Phys. } {\bf 2021}, {\em 18}, 21550121.
\bibitem{matondo}Matondo, D.; Maharaj, S.D. A Tolman-like Compact Model with Conformal Geometry. {\em Entropy} {\bf 2021}, {\em 23}, 1406.

\bibitem{rej}Bhar, P.;   Rej,  P.  Stable and self--consistent charged gravastar model within the framework of $f(R,T)$ gravity. {\em Eur. Phys. J. C } {\bf 2021}, {\em 81}, 763. 
\bibitem{RS}Sharma, R. Proper special conformal Killing vectors and the quadratic theory of gravity. {\em  J. Math. Phys.}, {\bf 1991}, {\em 32}, 1854.
\bibitem{MH2}Mak, M.K.; Harko, T. Can the galactic rotation curves be explained in brane world models? {\em Phy. Rev. D} {\bf 2004}, {\em 70}, 024010.
\bibitem{HM2} Harko, T.;  Mak, M.K.  Conformally symmetric vacuum solutions  of the gravitational field equations in the brane world model. {\em Ann. Phys.}, {\bf 2005}, {\em 319}, 471--492.
\bibitem{SIF}Sharif, M.;  Ismat Fatima, H. Static spherically symmetric solutions in $f(G)$  gravity. {\em Int. J. Mod. Phys. D} {\bf 2016}, {\em 25}, 1650083.
\bibitem{SHS}Sefiedgar, A.S.; Haghani, Z.;   Sepangi, H.R. Brane $f(R)$ gravity and dark matter. {\em  Phy. Rev. D}  {\bf 2012}, {\em 85}, 064012.
\bibitem{Bhar2}Bhar, P. Higher dimensional charged gravastar admitting conformal motion. {\em Astrophys. Space Sci.}  {\bf  2014}, {\em 354}, 457--462.
\bibitem{TD} Turkoglu, M.;  Dogru, M. Conformal cylindrically symmetric spacetimes in modified gravity. {\em  Mod. Phys. Lett. A} {\bf 2015}, {\em 30}, 1550202.
\bibitem{DRGR2}Das, A.; Rahaman, F.; Guha, B.K.; Ray,  S. Relativistic compact stars in $f(T)$ gravity admitting conformal motion.  {\em  Astrophys. Space Sci.} {\bf 2015}, {\em 358}, 36.
\bibitem{OS} Sert, O.  Radiation fluid stars in the non--minimally coupled $Y(R)F^2$ gravity. {\em  arXiv: 1611.03821v1} {\bf 2016}.
\bibitem{ZSRA}  Zubair, M.; Sardar,  L.H.; Rahaman, F.; Abbas, G. Interior solutions for fluid spheres in $f(R,T)$ gravity admitting conformal killing vectors.  {\em Astrophys. Space Sci.} {\bf 2016}, {\em 361}, 238.
\bibitem{DRGR}Das, A.; Rahaman, F.; Guha, B.K.;  Ray, S. Compact stars in $f(R, {\cal T})$ gravity. {\em Eur. Phys. J. C} {\bf 2016}, {\em 76}, 654.
\bibitem{SNN}Sharif, M.; Naz, S. Stable charged gravastar model in $f (R, T^2)$ gravity with conformal motion. {\em Eur. Phys. J. P.} {\bf 2022}, {\em 137}, 421.
\bibitem{BHL1} Bohmer, C.G.; Harko , T.; Lobo, F.S.N. Conformally traversable wormholes. {\em Phys. Rev. D}  {\bf 2007}, {\em 76}, 084014.
\bibitem{BHL} Bohmer, C.G.; Harko , T.; Lobo, F.S.N. Wormhole geometries with conformal motions.  {\em  Classical Quantum Grav.}  {\bf 2008}, {\em 25}, 075016.
\bibitem{RRKKK}Rahaman, F.; Ray, S.; Khadekar, G.; Kuhfittig, P.;  Karakar, I. {\em  Int. J. Theor. Phys.} {\bf 2015}, {\em 54}, 699.
\bibitem{K}Kuhfittig, P. Wormholes admitting conformal Killing vectors and supported by generalized Chaplygin gas. {\em Eur. Phys. J. C} {\bf 2015},  {\em 75}, 357.
\bibitem{SIF2}  Sharif, M.; Ismat Fatima, H. {Conformally symmetric traversable wormhole in $f(G)$ gravity. \em Gen. Relativ. Gravit.} {\bf 2016}, {\em 48}, 148.
\bibitem{kar} Kar,  S. Curious variant of the Bronnikov--Ellis spacetime.{\em  Phys. Rev. D} {\bf 2022}, {\em 105}, 024013.
\bibitem{sahoo} Mustafa, G.; Hassan, Z.; Sahoo, P.K.  Traversable wormhole inspired by non--commutative geometries in $f(Q)$ gravity with conformal symmetry. {\em  Ann. Phys.} {\bf 2022}, {\em 437}, 168751.
\bibitem{uref}Herrera, L; Di Prisco, A; Ospino, J. Non--static fluid spheres admitting a conformal Killing vector: Exact solutions.  {\em Universe} {\bf 2022} {\em  8}, 296.
\bibitem{epjc}Herrera, L.; Di Prisco , A.; Ospino, J.  Quasi--homologous evolution of self--gravitating systems  with vanishing complexity factor. {\em Eur. Phys. J. C} {\bf  2020}, {\em 80}, 631.
\bibitem{ps1} Herrera,  L. New definition of complexity for self--gravitating fluid distributions: The spherically symmetric case.  {\em Phys. Rev. D} {\bf  2018}, {\em 97}, 044010.
\bibitem{ps2}Herrera, L.; Di Prisco , A.; Ospino, J. Definition of complexity for dynamical spherically symmetric dissipative self--gravitating fluid distributions. {\em  Phys. Rev. D} { \bf  2018}, {\em 98}, 104059.
\bibitem{sc} Herrera, L.;  Ospino,  J.;  Di Prisco, A.; Fuenmayor, E.; Troconis, O. Structure and evolution of self--gravitating objects and the orthogonal splitting of the Riemann tensor. {\em  Phys. Rev D} {\bf 2009}, {\em 79}, 064025.

\bibitem{chan1}Chan, R. Collapse of a radiating star with shear. {\em Mon. Not. R. Astron. Soc.} {\bf 1997}, {\em 288}, 589--595.
\bibitem{isjc}Israel, W. Singular hypersurfaces and thin shells in general relativity. {\em Il Nuovo C. B} {\bf 1966}, {\em 44}, 1?14.
\bibitem{19nt}  Israel,  W.  Nonstationary irreversible thermodynamics: A causal relativistic theory.  {\em  Ann. Phys.} (NY) {\bf 1976}, {\em 100}, 310--331.
\bibitem{20nt} Israel, W.;  Stewart, J. Thermodynamic of nonstationary and transient effects in a relativistic gas. {\em  Phys. Lett. A} {\bf 1976}, {\em 58}, 213--215.
\bibitem{21nt}Israel, W.;  Stewart, J. Transient relativistic thermodynamics and kinetic theory. {\em Ann. Phys.} (NY) {\bf 1979}, {\em 118},  341--372.
\bibitem{t8}Triginer, J.;  Pavon, D. On the thermodynamics of tilted and collisionless gases in Friedmann--Robertson--Walker spacetimes. {\em  Class. Quantum Grav.} {\bf  1995}, {\em 12}, 199.
\bibitem{20n}Schwarzschild, M.  \textit{Structure and Evolution of the Stars};  Dover:  New York, USA, 1958.
\bibitem{21n}Kippenhahn , R.; Weigert,  A. \textit{ Stellar Structure and Evolution};  Springer Verlag:  Berlin, Germany, 1990.
\bibitem{22n} Hansen, C;  Kawaler,  S. \textit{Stellar Interiors: Physical  Principles, Structure and Evolution};  Springer Verlag:  Berlin, Germany, 1994.


\end{thebibliography}
\end{document}